\documentclass[11pt,twoside]{article}

\usepackage{asp2006}
\usepackage{epsf}

\begin{document}
\title{Properties of bars in the local universe}   
\author{J. M\'endez-Abreu}   
\affil{INAF-Osservatorio Astronomico di Padova, Padova, Italy 
\author{\rm J. A. L. Aguerri}
\affil{Instituto de Astrof\'isica de Canarias, La Laguna, Spain}
\author{\rm E. M. Corsini}
\affil{Dipartimento di Astronomia, Universit\`a di Padova, Padova, Italy}
}


\begin{abstract}
We studied the fraction and properties of bars in a sample of about
3000 galaxies extracted from SDSS-DR5. This represents a volume
limited sample with galaxies located between redshift $0.01<z<0.04$,
absolute magnitude $M_{r}>-20$, and inclination $i <
60^{\circ}$. Interacting galaxies were excluded from the sample. The
fraction of barred galaxies in our sample is $45\%$. We found that
$32\%$ of S0s, $55\%$ of early-type spirals, and $52\%$ of late-type
spirals are barred galaxies. The bars in S0s galaxies are weaker than
those in later-type galaxies. The bar length and galaxy size are
correlated, being larger bars located in larger galaxies. Neither the
bar strength nor bar length correlate with the local galaxy
density. On the contrary, the bar properties correlate with the
properties of their host galaxies. Galaxies with higher central light
concentration host less and weaker bars.
\end{abstract}

\section{Detection and characterization of bars}

Large galaxy surveys allows to study the bar properties in the local
universe for a statistically significant sample of galaxies. We
selected in SDSS-DR5 a volume-limited sample of about 3000 galaxies
with redshift $0.01<z<0.04$, absolute magnitude $M_{r}>-20$, and
inclination $i < 60^{\circ}$. They were selected to no show evidence
of interaction.
Bars were detected using two different methods: ellipse fitting and
Fourier analysis. They were calibrated and tested using mock barred
and non-barred galaxies similar to the observed ones. We present here
the results obtained with the ellipse fitting method, since it
resulted to be more efficient than Fourier analysis in detecting bars.
We found a bar in $45\%$ of our disk galaxies, in good agreement with
Barazza et al. (2007). As far as the morphological type concerns,
$32\%$ of S0s, $55\%$ of early-type spirals, and $52\%$ of late-type
spirals are barred galaxies.  The bar length was measured either as
the radius corresponding to the maximum isophotal ellipticity or as
the radius where the isophotal position angle changes more than
$5^{\circ}$. It correlates with the galaxy size, which was defined as
$2.5 \times r_{90}$ where $r_{90}$ is the radius enclosing $90\%$ of
the total flux. The bar strength was calculated using the parameter
$f_{bar}$ (Whyte et al. 2002). Bars in S0s galaxies are weaker than
those in later-type galaxies.

\begin{figure}
   \plotone{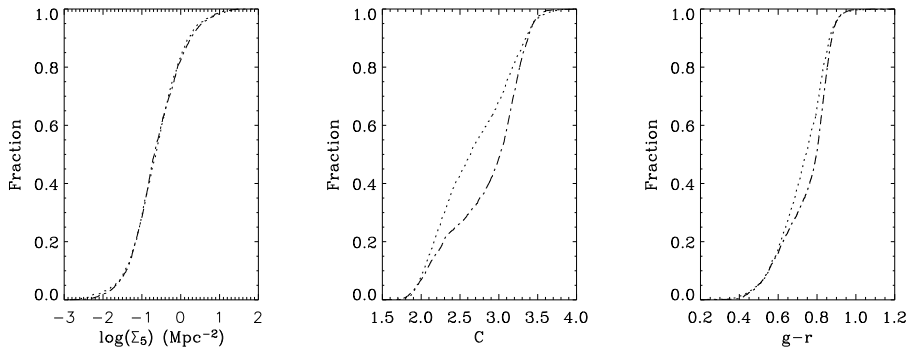}
    \caption{Cumulative fraction of barred (dotted line) and non-barred
      (dashed-dotted line) as function of the local density (left
      panel), light concentration (central panel) and $g-r$ galaxy
      color of the galaxies (right panel).}
   \label{fig:cum}
   \end{figure}

\section{Internal versus external processes}
  
The local galaxy density $\Sigma_{5}$ was estimated using the method
of the distance to the fifth nearest neighbor. There is no correlation
between the presence (Fig. \ref{fig:cum}, left panel), strength, and
length of a bar with $\Sigma_{5}$. We also investigated the relation
of the different bar properties with the light concentration index
$C=r_{90}/r_{50}$. We found that bars were preferentially hosted in
galaxies with lower light concentration (Fig. \ref{fig:cum}, central
panel). Moreover, bars are preferentially hosted by bluer galaxies
(Fig. \ref{fig:cum}, right panel)

\section{Conclusions}

Bars were observed in $45\%$ of our disk galaxies, being more common
in early- and late-type spirals than in S0s galaxies. We found that
larger bars are located in larger galaxies, indicating the interplay
between the bar and disk components. No difference between the local
galaxy density of barred and non-barred galaxies was found, in
addition, neither the length nor strength of the bars are correlated
with the local environment. These results suggest that local
environment does not play an important role in bar formation and
evolution. A statistical significant difference between the central
light concentration of barred and non-barred galaxies was found. The
former were located in less concentrated galaxies. This difference
could explain the lower fraction and the weakness of bars detected in
S0 galaxies in comparison with the other types.


\end{document}